\documentclass[preprint,prb,superscriptaddress]{revtex4-2}   
\usepackage[utf8]{inputenc}
\usepackage{graphicx}% Include figure files 

\begin{document}  
\title{Spontaneous symmetry breaking in plasmon lattice lasers}
\author{Nelson de Gaay Fortman}
    \affiliation{Institute of Physics, University of Amsterdam, NL1098XH Amsterdam, The Netherlands}  
    \affiliation{Department of Physics of Information in Matter and Center for Nanophotonics, NWO-I Institute AMOLF, Science Park 104, NL1098XG Amsterdam, The Netherlands}
   \author{Radoslaw Kolkowski}
    \affiliation{Department of Physics of Information in Matter and Center for Nanophotonics, NWO-I Institute AMOLF, Science Park 104, NL1098XG Amsterdam, The Netherlands}
    \affiliation{Department of Applied Physics, Aalto University, P.O. Box 13500, FI-00076 Aalto, Finland}
\author{Debapriya Pal}
    \affiliation{Department of Physics of Information in Matter and Center for Nanophotonics, NWO-I Institute AMOLF, Science Park 104, NL1098XG Amsterdam, The Netherlands}
\author{Said R. K. Rodriguez}
    \affiliation{Department of Physics of Information in Matter and Center for Nanophotonics, NWO-I Institute AMOLF, Science Park 104, NL1098XG Amsterdam, The Netherlands}
\author{Peter Schall}
    \affiliation{Institute of Physics, University of Amsterdam, NL1098XH Amsterdam, The Netherlands}
\author{A. Femius Koenderink}
    \affiliation{Department of Physics of Information in Matter and Center for Nanophotonics, NWO-I Institute AMOLF, Science Park 104, NL1098XG Amsterdam, The Netherlands}
    \affiliation{Institute of Physics, University of Amsterdam, NL1098XH Amsterdam, The Netherlands}
\email{f.koenderink@amolf.nl}

\date{\today}

\begin{abstract}
Spontaneous symmetry breaking (SSB) is key for our understanding of phase transitions and the spontaneous emergence of order. Photonics provide versatile systems to study SSB. In this work, we report that for a two-dimensional (2D) periodic nonlocal metasurface with gain, SSB occurs in the lasing transition, breaking parity symmetry. We study diffractive hexagonal plasmon nanoparticle lattices, where the $\mathit{K}$-points in momentum space provide two modes that are exactly degenerate in frequency and identically distributed in space. Using femtosecond pulses to energize the gain medium, we simultaneously capture single-shot real-space and wavevector resolved ‘Fourier’ images of laser emission. By combining Fourier- and real-space, we resolve the two order parameters for which symmetry breaking simultaneously occurs: spatial parity and $U(1)$ (rotational) symmetry breaking, evident respectively as random relative mode amplitude and phase. Thereby, we quantify for the first time SSB in 2D periodic metasurfaces. These currently receive much interest as experimentally accessible implementations of seminal solid-state physics Hamiltonians and provide a large design space for exploring SSB in scenarios with different symmetries, mode degeneracies and topological properties. The methodology reported in this work is generally applicable to 2D plasmonic and dielectric metasurfaces and opens numerous opportunities for the study of SSB and emergence of spatial coherence in metaphotonics.
\end{abstract}
\maketitle

In nature, there exist many physical systems that can spontaneously and abruptly evolve into an asymmetric state. This phenomenon of spontaneous symmetry breaking (SSB), is observed for instance in  phase transitions, such as in the emergence of ferromagnetism, superfluidity, superconductivity, and Bose-Einstein condensation. In photonics, spontaneous symmetry breaking occurs for instance at the lasing transition, marked by the emergence of a randomly picked phase, and in degenerate resonant systems imbued with gain or nonlinearity, where SSB expresses as a random imbalance in the population of the modes. SSB has recently been studied in evanescently coupled `photonic molecule’ nanocavities \cite{hamel_spontaneous_2015, garbin_spontaneous_2022}, in microcavities with degeneracy between counter-propagating modes (whispering gallery mode systems), with either gain or a Kerr nonlinearity \cite{m-tehrani_mode_1977, sorel_unidirectional_2002, zhukovsky_bistability_2009,  cao_experimental_2017, del_bino_symmetry_2017, keitel_single-pulse_2021}, and in exciton-polariton microcavities and microcavity lattices \cite{wertz_spontaneous_2010, ohadi_spontaneous_2012, sala_stochastic_2016, sigurdsson_spontaneous_2019}. There is a large interest in understanding and harnessing the mechanisms of SSB in photonics. Firstly, because photonic systems are ideal to study SSB as a general phenomenon, as nonlinear photonic resonators allow to synthesize many dynamical nonlinear differential equations with SSB behavior, and in photonics one can measure system response with high precision and over many decades in time. Secondly, from the application viewpoint, SSB is envisioned as key to realizing ultra-small flip-flop optical memories, optically controllable circulators and isolators, and all-optical switching \cite{del_bino_symmetry_2017, haelterman_all-optical_1991, maes_switching_2006}.

Plasmon lattice lasers are periodically arranged lattices of plasmon nanoparticles embedded in a gain waveguide slab, and were first reported in 2003 \cite{stehr_low_2003}, and have since then been studied extensively \cite{suh_plasmonic_2012, wang_ultrafast_2019, hoang_millimeter-scale_2017, winkler_dual-wavelength_2020, hakala_lasing_2017, guo_lasing_2019, heilmann_quasi-bic_2022, schokker_lasing_2014, schokker_statistics_2015, schokker_systematic_2017, guo_spatial_2019}. They are akin to distributed feedback (DFB) lasers \cite{kogelnik_stimulated_1971}, but the weak feedback mechanism of a dielectric Bragg grating is replaced by nonlocal diffractive  plasmon modes. These so-called plasmon surface lattice resonances have high quality factor and strong plasmonic near fields. These unique properties provide strong feedback \cite{schokker_lasing_2014}, exceptional robustness to disorder \cite{schokker_statistics_2015}, and ultrafast gain dynamics \cite{wang_ultrafast_2019}. In the plasmon lattice system, one can tailor the unit cell resonance (controlled by antenna size and shape) and lattice symmetry at will, providing opportunities to study a rich family of band structure physics, such as honeycomb and Kagom\'{e} lattices with topological properties \cite{abass_tailoring_2014, weick_dirac-like_2013, saito_valley-polarized_2021}, bound states in the continuum modes \cite{trinh_coexistence_2022}, and exceptional points \cite{kolkowski_pseudochirality_2021}.  

In this work, we demonstrate spontaneous symmetry breaking in hexagonal plasmon lattice lasers, using the intrinsic degeneracy of Bloch modes at the $\mathit{K}$-symmetry points in reciprocal space, which are defined as the corners of the first Brillouin zone. Lasing from these points has been reported before in photonic crystal and plasmonic lattice lasers~\cite{notomi_directional_2001,guo_lasing_2019}. However, these works did not report spontaneous symmetry breaking.  We developed a new methodology where we use a pulsed femtosecond laser as pump to bring the system to lasing in every single-shot, and we simultaneously perform real-space and Fourier-space microscopy to map the relative intensity and phase of the lasing modes from shot to shot. We uncover that  this system at the same time shows parity symmetry breaking, observable in the direction of light emission, and rotational SSB, i.e. $U(1)$ symmetry breaking, observable as a random choice of relative phase between the lasing modes \cite{gartner_spontaneous_2019}. While conventionally studied coupled microcavity systems \cite{hamel_spontaneous_2015, sala_stochastic_2016, garbin_spontaneous_2022, wertz_spontaneous_2010, ohadi_spontaneous_2012, sigurdsson_spontaneous_2019} are ideal for two-mode coupling or optical simulators of nonlinear Hamiltonians with just nearest-neighbor interactions, metasurfaces provide an even richer design space to design symmetries, mode degeneracies, and long range interactions. The extended 2D nature of the lattice laser also means that one can go beyond mapping only overall mode populations: our microscopy methods resolve spatial structure in the SSB, giving direct insight in the spatial structure of the spontaneously emerging coherence. Therefore, this work opens a rich venue for studying spontaneous symmetry breaking in nanophotonic metasurfaces. 
 
\begin{figure*}[!ht]
    \centering
    \includegraphics[width=0.75\textwidth]{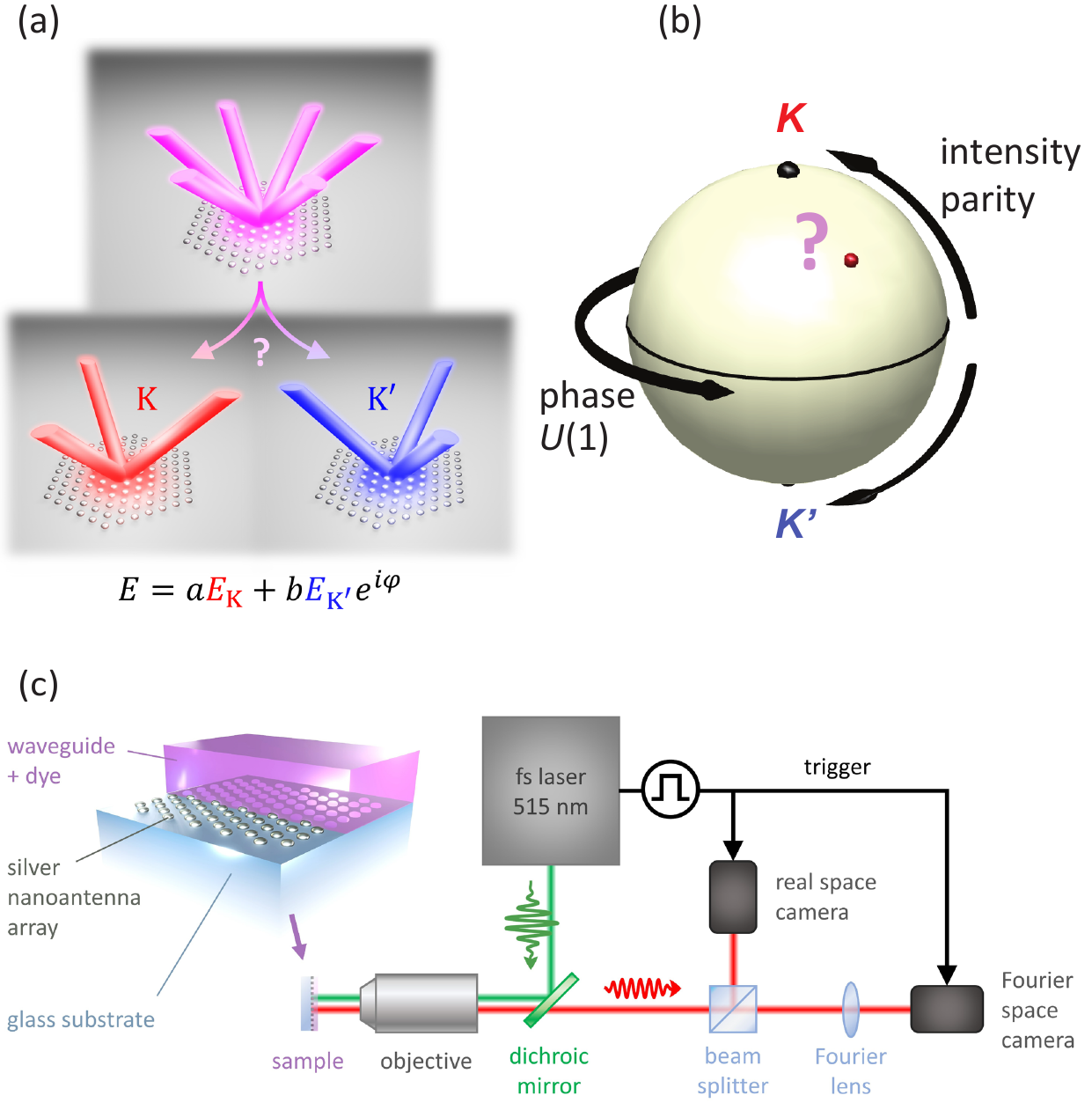}
    \caption{\textbf{Single-shot microscopy retrieves relative intensity and phase of symmetry broken $\mathit{K}$ and $\mathit{K'}$ laser modes.} Panel (a): $\mathit{K}$-point lasing in hexagonal plasmon lattice lasers occurs on two decoupled modes, degenerate in frequency and space, and only differing in parity. Spontaneous symmetry breaking occurs in relative amplitude between the $\mathit{K}$ and $\mathit{K'}$-mode (parity breaking), and in relative phase ($U(1)$ symmetry breaking). The phase space maps to the unit sphere (b), where the distance from the equator maps parity-breaking, and the azimuth maps relative phase. Panel (c): plasmon lattices are embedded in a planar polymer waveguide with organic dye to provide gain.  We study lasing in a high-NA microscope with single-shot real-space and Fourier imaging capabilities, synchronized to a 20 Hz train of pump pulses (515 nm, 250 fs).}
\end{figure*}

\begin{figure*}[!ht]
    \centering
    \includegraphics[width=0.65\textwidth]{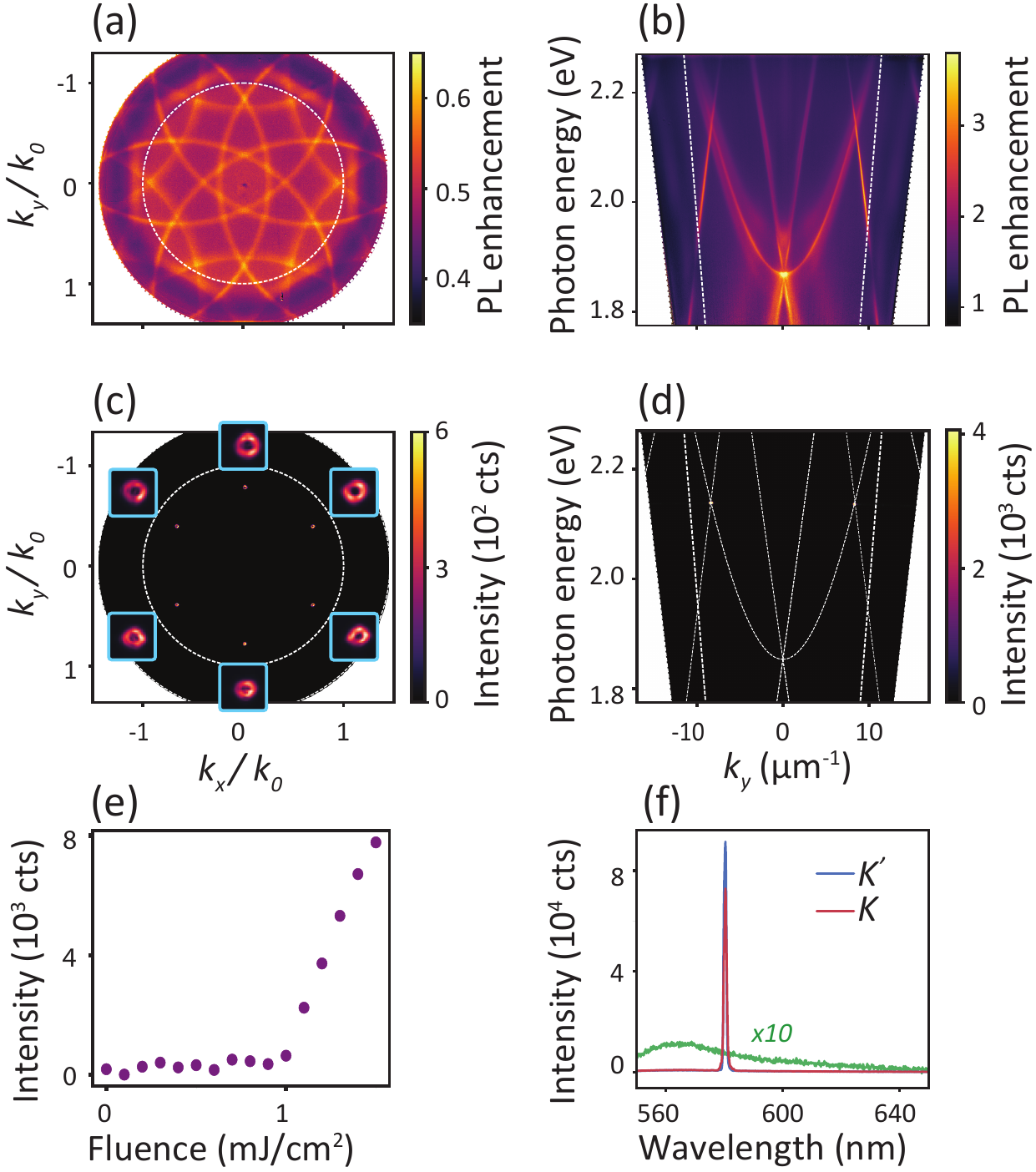}
    \caption{\textbf{$\mathit{K}$-point lasing averaged over multiple laser shots.} Panel (a): Fourier image of photoluminescence (PL) shows the isofrequency contours of the guided mode dispersion curves that are repeated in the momentum space due to periodicity (below lasing threshold). Threeway crossings just within the NA $=1$ circle (dashed) correspond to the $\mathit{K}$-points. Panel (b): below threshold PL band structure showing $\mathit{K}$-point intersections at $\hbar\omega=2.15$ eV. Panel (c): above-threshold Fourier image showing lasing at the six $\mathit{K}$-points in Fourier-space, with 5$\times$ enlarged beam spots as insets (averaged over 150 shots). The beam spots are doughnut-shaped due to the quasi-BIC nature of the lasing condition. Panel (d): lasing at the $\mathit{K}$-points in the dispersion image (six shots summed), with calculated free photon lines. Panel (e): input-output power curve. Panel (f): above threshold, the broad fluorescence spectrum (green curve) narrows to two sharp lasing peaks. The pump fluence of the lasing spectra is 3$\times$ larger than the threshold fluence. The emission wavelengths at $\mathit{K}$- and $\mathit{K'}$-point are identical.}
\end{figure*}

Fig 1 illustrates our approach. We study hexagonal periodic lattices of plasmonic nanodisks (Ag, diameter 80 nm and height 30 nm) embedded in a polymer waveguide doped with a laser dye (fabraction described in Methods). This layer acts both as  gain medium and as planar waveguide with a single transverse electric (TE) and transverse magnetic (TM) mode. The dominantly in-plane nanoantenna polarizability makes this geometry especially favorable to TE-mode distributed feedback lasing. The 500~nm lattice pitch creates a $\mathit{K}$-point lasing condition near 580 nm wavelength. The 2D hexagonal lattice provides degeneracy at the K-point both in frequency and space. The Brillouin zone has six $\mathit{K}$-points that fall apart in two decoupled sets (henceforth $\mathit{K}$ and $\mathit{K'}$ points), each consisting of three points that are internally connected by a reciprocal lattice vector. In a scalar description, at a given $\mathit{K}$-point, there exist three modes, two of which form a pair of doubly degenerate $E$-modes and the remaining $A_1$ mode is the only mode that forms a flat band \cite{sakoda_optical_2005} required for lasing. While for polarized waves, group theory is more involved, for the first order TE waveguide mode the same separation into a degenerate doublet and one flat band holds \cite{ochiai_dispersion_2001}. The $\mathit{K}$ and $\mathit{K'}$ flat band modes are not only frequency-degenerate  but also their local fields $\mathbf{E}_{K,K'}(\mathbf{r})$ are  exactly identical in terms of energy density ($|\mathbf{E}_{K}(\mathbf{r})|^2=|\mathbf{E}_{K'}(\mathbf{r})|^2$). The only difference is in the mirrored wavevector content. This offers the condition for spontaneous parity symmetry breaking: lasing just on the  $\mathit{K}$, or just on the $\mathit{K'}$ condition will correspond to three as opposed to six far-field output spots, with mirrored orientation (Fig 1(a)). Lasing of both modes in some superposition will furthermore imply the spontaneous emergence of a random relative phase $\varphi$, breaking $U(1)$ symmetry \cite{graham_laserlight_1970}. In this work, we visualize the phase space for such SSB on the unit sphere (Fig 1(b)): moving away from the equator represents parity breaking (pure $\mathit{K}$ resp  $\mathit{K'}$ lasing at the north and south pole), while azimuth represents the relative phase.

To map the symmetry breaking both in intensity parity and in phase, we developed a simultaneous single-shot real- and Fourier-space microscopy method (Fig 1(c)). The samples are pumped by 250 fs pulses at 515 nm  wavelength. Emission collected by a high NA microscope objective is split in two optical tracks, each with a camera synchronized with the 20 Hz laser pulse train. By imaging real-space in one track, while inserting a Fourier lens in the other, we obtain statistics to correlate Fourier-space and real-space output over long sequences of single-shot experiments. To assess the dispersion relation underlying the laser behavior, we collect photoluminescence (PL) enhancement Fourier and dispersion images (Figs 2(a, b)) by pumping at low fluence yet high repetition rate (1 MHz). Refer to Methods for more details. The PL enhancement Fourier image (Fig 2(a)) shows that exactly three circular bands cross at each $\mathit{K}$-point. These are the 2D slab waveguide modes folded by the lattice periodicity \cite{vaskin_light-emitting_2019}. The intersections are also clearly visible in the dispersion (Fig 2(b)). The three-fold denegeracy is consistent with the fact that $\mathit{K}$-point lasing corresponds to feedback on three reciprocal lattice vectors $G$ that form a closed triangular loop~\cite{notomi_directional_2001}. The dispersion image, taken by projecting the $k_x=0$ slice of the Fourier image on the entrance slit of a spectrometer, samples one of the $\mathit{K'}$ points (negative $k_y$), and one of the $\mathit{K}$ points (positive $k_y$), at an energy of 2.15 eV.

To measure lasing, we operate in single-shot mode. Increasing the pump pulse energy leads to a sudden nonlinear increase in output power at the $\mathit{K}$-points  together with a very clear spectral narrowing (Figs 2(e,f)). The thresholds of 1 mJ/cm$^2$ (50 nJ pulse energy) are consistent with earlier works on plasmon lattice lasing \cite{schokker_lasing_2014, guo_lasing_2019, guo_spatial_2019}. Fourier images integrated over many shots (Fig 2(c)) distinctly show six high-intensity spots at the $\mathit{K}$-points. Each has a doughnut shape, reflecting the fact that the lasing mode is prevented from radiating directly into the far-field due to its symmetry. This effect is known as a symmetry-protected optical bound state in the continuum (BIC). Instead, the emission is possible from the slightly non-symmetric modes around the lasing modes (the so-called quasi-BICs), to which the emitted light is scattered \cite{imada_multidirectionally_2002, zhang_observation_2018}. The dispersion image for lasing is shown in Fig 2(d). Two intensity peaks from lasing, at the $\mathit{K}$ and $\mathit{K'}$ point, occur at the same wavelength (Fig 2(f)), highlighting the frequency degeneracy that the hexagonal lattice symmetry guarantees for $\mathit{K}$/$\mathit{K'}$ modes.

\begin{figure*}[!ht]
    \centering
    \includegraphics[width=0.65\textwidth]{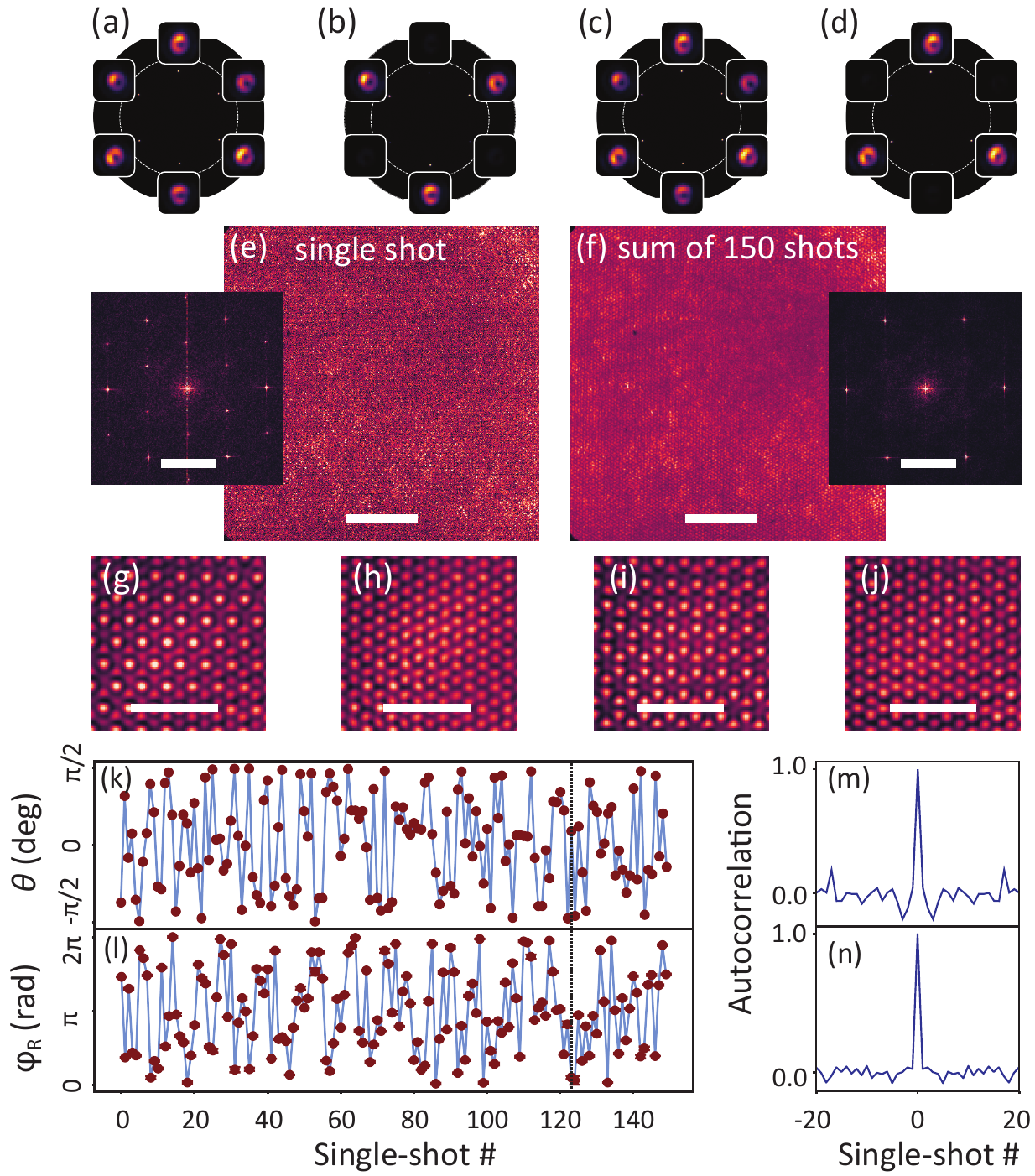}
    \caption{\textbf{Single-shot $\mathit{K}$-point lasing and SSB in Fourier- and real-space.} Panels (a-d): single-shot Fourier images taken from a single experiment run. Shots 1-3 (panels a-c) are directly subsequent to each other. Panels (b,d) show parity breaking towards pure $\mathit{K}$/$\mathit{K'}$-point lasing, while panels (a,c) show cases of equal superposition. Panel (e): single-shot and ensemble-averaged real-space images (scale bar: 10 $\mu$m). Insets show 2D Fourier transforms (scale bar: 10 $\mu$m$^{-1}$). Panels (g-j): Fourier-filtered closeups of the real-space images (scale bar: 3 $\mu$m), corresponding to the Fourier images (a-b).  Panels (k,j): sequence of relative amplitude (parity-breaking) and phase ($U(1)$ symmetry breaking) for a 150-shot data sequence. The amplitude contrast is plotted as  angle $\theta$ on the unit sphere defined through $\tan \theta/2=(I_K-I_{K'})/(I_K+I_{K'})$. Panels (m,n): temporal autocorrelations of the data in (k,l). The traces are taken at 1.8 mJ/cm$^{2}$ pump power. The dashed line highlights the location of shots (a,b,c).}
\end{figure*}

Spontaneous symmetry breaking in parity is directly evident in sequences of single-shot Fourier images. Figs 3(a-d) show four single-shot images taken from a run of several hundreds of laser shots. For each of the six $\mathit{K}$-point lasing spots, we observe very large shot-to-shot differences in intensity. Figs 3(a,b,c) show three directly subsequent shots, in the first of which both $\mathit{K}$ and $\mathit{K'}$-points appear equally bright, while the next frame has hardly any $\mathit{K'}$-point emission (only three lasing spots at the $\mathit{K}$ points). The opposite parity breaking entirely towards $\mathit{K'}$-lasing is evident in panel (d). In the entire data sequence, no spot patterns occur \textit{except} linear superpositions of the $\mathit{K}$ and $\mathit{K'}$-triplets. Given the observation that the lasing output is intermittent with a different intensity ratio of $\mathit{K}$ and $\mathit{K'}$ modes in each shot, a relevant question is what the relative \textit{phase} between the two modes is in a given realization, and how that phase is statistically distributed over all shots. Since in the Fourier-space, the $\mathit{K}$ and $\mathit{K'}$ modes have no overlap,  Fourier imaging can not reveal relative phase. However, the relative phase can be determined from single-shot real-space images, which we acquire simultaneously with Fourier images. Upon crossing the lasing threshold, the real-space appearance of the sample transitions from the diffuse glow of fluorescence to a structured speckled pattern, that directly evidences the emergence of both spatial and temporal coherence \cite{schokker_statistics_2015}. Fig 3(e) shows a typical single-shot real-space image over an area of $44 \times 44$ $\mu$m$^2$,  while Fig 3(f) shows the sum over many laser shots.  In both results, the hexagonal lattice symmetry is visible. However, 2D Fourier transforms (insets) reveal a marked difference. The ensemble-averaged image \textit{only} shows Fourier components commensurate with the lattice of particles (six sharp peaks in the Fourier transformed images). The single-shot image instead shows additional spatial structure, evident as a 6-tuple of Fourier peaks at wave vectors \textit{shorter} by a factor $\sqrt{3}$. This points at the emergence of \textit{longer} range periodicities. Motivated by this observation, we apply Fourier-domain filtering (explained in supplement) and examine closeups of Fourier-filtered real-space data. Figs 3(g-j) correspond directly to the single-shot Fourier images in Figs 3(a-d).  Whenever only $\mathit{K}$ (shot 2) or only $\mathit{K'}$ (shot 4) lasing occurs, the intensity patterns  are simply hexagonal with periodicity identical to the particle lattice. In stark contrast, when both $\mathit{K}$- and $\mathit{K'}$-points substantially lase (shots 1 and 3), we observe a honeycomb supercell pattern. Moreover, while shots 1 and 3 have very similar Fourier-images, the spatial patterns are distinct: the shot 3 honeycomb pattern is clearly shifted horizontally and vertically compared to shot 1. This variation in spatial pattern encodes the random relative phase between $\mathit{K}$- and $\mathit{K'}$-point lasing. 

To explain how we retrieve relative phase from real-space images, we consider a simple scalar coupled-mode model for $\mathit{K}$/$\mathit{K'}$-point Bloch modes \cite{wheeldon_symmetry_2007, malterre_symmetry_2011}. The $A_1$ mode at the $\mathit{K}$-point is a coherent superposition of three plane waves of identical amplitude and phase, with in-plane momentum given by the three $\mathit{K}$-points:

\begin{equation}
    \mathbf{E}_{\mathit{K}} = e^{i\mathbf{\mathit{K}_1}\cdot\mathbf{r}} + e^{i\mathbf{\mathit{K}_2}\cdot\mathbf{r}} + e^{i\mathbf{\mathit{K}_3}\cdot\mathbf{r}}
\end{equation}

The $\mathit{K'}$ field is described by the same equation but substituting  the $\mathit{K'}$-points as in-plane momenta. For both of these modes individually, the local intensity distributions correspond to identical simple hexagonal patterns with peak intensities at the lattice positions. However, for superpositions $\mathbf{E}_{T} = a \mathbf{E}_{\mathit{K}} + b \mathbf{E}_{\mathit{K'}}e^{i\varphi_R}$ the intensity distributions acquire  a superlattice periodicity that is $\sqrt{3}$ times larger than the original lattice pitch. The relative amplitudes $a$ and $b$ affect the contrast, but not the topology of these honeycomb intensity patterns. The alignment of the superlattice intensity pattern relative to the underlying particle lattice depends on the relative phase $\varphi_R$ (see supplement for a plot catalogue). The simple coupled-mode model qualitatively rationalizes all our observations, namely: (i) whenever  lasing is purely on the $\mathit{K}$ or $\mathit{K'}$-point (shots 2 and 3), the intensity distribution is simply hexagonal. The real-space intensity distributions are identical, even though the far-field directions of lasing emission are mirrored; (ii) honeycomb patterns occur when $\mathit{K}$ and $\mathit{K'}$ modes both lase, and the observed spatial shift from frame to frame can be understood as shot-to-shot variations in relative phase $\varphi_R$ (shots 1 and 4 in Fig 3); (iii)  the incoherent sum over all frames washes out this interference and is simply hexagonal (Fig 3(f)). To extract the relative phase $\varphi_R$ for each laser shot, we take a small section of the real-space data (as selected for Figs 3(g-j)) and fit the coupled mode model, where the relative amplitude is not a free parameter, but fixed by the  Fourier images (see supplement for methodology).

Figs 3(k,l) report on the extracted fluctuations in parity-breaking ($\mathit{K}$/$\mathit{K'}$ contrast)  and relative phase (random phase) in a typical measurement run. In line with Fig 1(b), we quantify the $\mathit{K}$/$\mathit{K'}$ intensity contrast as an angle $\theta$ defined for intensity $I_K$ in the $\mathit{K}$-modes and $I_{K'}$ in the $\mathit{K'}$-modes as $\tan \theta/2=(I_K-I_{K'})/(I_K+I_{K'})$. From the time traces we conclude that both the intensity ratio and relative phase are picked randomly and independently from shot to shot. The normalized temporal autocorrelations (panels m,n) for both the $\mathit{K}$/$\mathit{K'}$-intensity ratio and the relative phase $\varphi_R$ show no correlation beyond $\Delta n=0$. This observation is consistent with the notion that the symmetry breaking is not from any \textit{explicit} symmetry breaking, but is purely \textit{spontaneous}. In the system at hand, the underlying modes are by construction degenerate in both space and frequency. This implies that the two modes are pumped equally, regardless of the spatial non-uniformity resulting, e.g., from fabrication imperfections  or from the spatial variation of the pump intensity. For instance, in our experiment we were unable to impart explicit symmetry breaking with linear pump polarization, even though the pump field does have near field hot spots that rotate with input polarization.  To conclude, this system shows pure spontaneous symmetry breaking, which is a result of its inherent protection from explicit symmetry breaking caused by an inhomogeneous gain distribution. Therefore, we argue that this system stands out as a platform for exploring SSB in the lasing transition, as compared with systems that do not enjoy this protection (for instance, coupled cavity lasers \cite{hamel_spontaneous_2015}).

To visualize how the SSB order parameters of this plasmon lattice laser sample phase space, we consider the distribution of the intensity contrast parameter $\theta$ and the phase $\varphi_R$ in Fig 4. For a run of 900 shots, we place each shot on the unit sphere in Fig 4(a), while Fig 4(b) and (c) report histogrammed results for each parameter individually.  As regards the parity breaking, we construct a histogram by a projection method that not only takes into account the $\mathit{K}$/$\mathit{K'}$ contrast, but that also reports on whether spot patterns are strict superpositions of just $\mathit{K}$/$\mathit{K'}$-lasing (see Methods). In short, Fig 4(b) reads as follows: if all six lasing spots are equally bright, then $c_+=1$ and $c_-=0$. If instead lasing emission is just in the $\mathit{K}$ (resp. $\mathit{K'}$) directions, then $c_+=\pm c_-=\sqrt{2}$. Any other linear combination of pure $\mathit{K}$ and  $\mathit{K}'$-point lasing will result in a plot coordinate on the circular arc connecting these extremes and at angle $\theta/2$ relative to the $x$-axis. Finally, any hypothetical intensity pattern that is \textit{not} a linear superposition of $\mathit{K}$ and $\mathit{K'}$-point lasing (e.g., if just a single laser spot would appear) results in a coordinate interior to the circular arc. The histogram  evidences that no such patterns occur, verifying that all laser shots are \textit{bona fide} linear combinations of $\mathit{K}$ and $\mathit{K'}$-point lasing. The variation in contrast $\theta$ essentially samples the entire interval from $-\pi/2$ to $\pi/2$ uniformly. Analysis as function of  pump power  (see supplement) shows that the histogram is identical irrespective of how high above threshold one operates. 

This expression of symmetry breaking is comparable to recent results on orbital angular momentum microlasers \cite{keitel_single-pulse_2021}, and different from the more usually reported bistability in which a system switches randomly  to just one of the  initially degenerate modes \cite{hamel_spontaneous_2015, garbin_spontaneous_2022}.  One can ask if this difference arises from the pulsed nature of our experiment, as previous studies on bistability either used quasi-CW conditions, or pulses substantially longer than the intrinsic gain dynamics. To verify if the pumping conditions affect the observed SSB character, we repeated the experiments with 500 ps instead of 250 fs pulses. These pulses are longer  than the picosecond time scales associated with distributed feedback lase dynamics. Despite pumping the system with such long pulses, we observe the same continuous histogram. Furthermore,
the fact that the real-space maps encode the relative phase rules out the possibility of rapid switching between the $\mathit{K}$ and $\mathit{K'}$ lasing within a single shot. Turning to the phase, Fig 4(c) plots the probability density for observing ($\cos\varphi_R,\sin\varphi_R)$. The relative phase between the $\mathit{K}$ and $\mathit{K'}$ is picked randomly as expected for $U(1)$ symmetry breaking, where a system's phase evolves as a second order phase transition from a highly symmetric (spontaneous emission) to locking in a single phase above lasing threshold \cite{graham_laserlight_1970, gartner_spontaneous_2019}.  Finally, we note that breaking the parity symmetry and $U(1)$ symmetry lead to uncorrelated choices in $\theta$ and $\varphi_R$ on the unit sphere.

\begin{figure*}[!ht]
    \centering
    \includegraphics[width=1\textwidth]{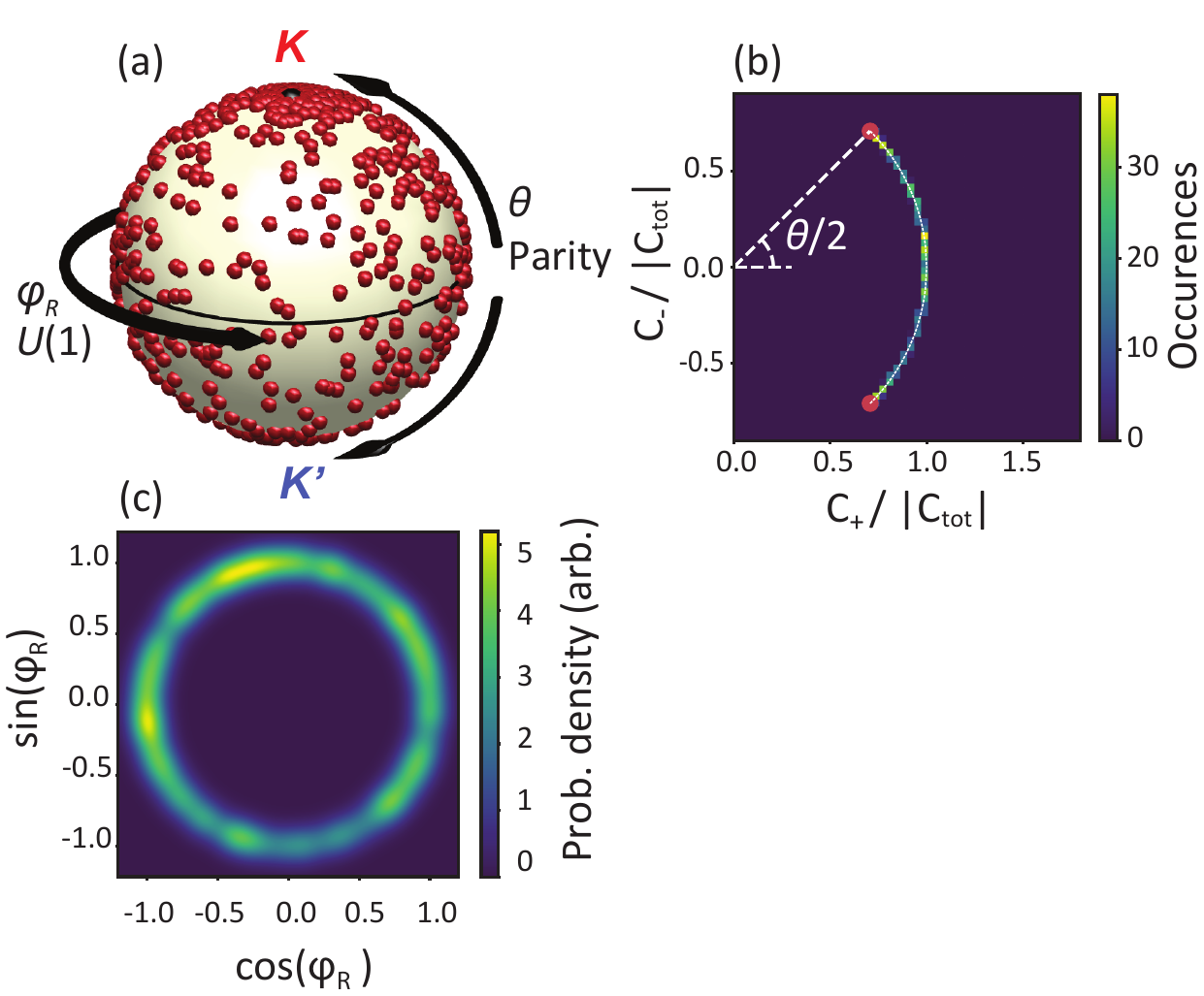}
    \caption{\textbf{SSB order parameters $\theta$ and $\varphi_R$, mapped onto the surface of a unit sphere.} Panel (a): SSB phase space samples in a 900-shot measurement run. Each shot is plotted as a data point on the unit sphere according to its relative phase (azimuth $\varphi_R$) and relative amplitude (parity breaking, distance $\theta$ from equator).  Panels (b,c): histograms of the parity breaking $\theta$ and random phase $\varphi_R$. }
\end{figure*}

%\section{Conclusion and outlook}
In summary, we  observed spontaneous symmetry breaking in $\mathit{K}$-point plasmon lattice lasers, where the symmetry is broken both in relative amplitude and phase between the two ($\mathit{K}$ and $\mathit{K'}$) lasing modes. The SSB in intensity implies a breaking of parity symmetry, while the SSB in phase is a $U(1)$ symmetry breaking. A key aspect of our work is the methodology of simultaneous single-shot Fourier- and real-space imaging, which should be contrasted to the earlier time-averaged measurements. This methodology is generally applicable to a large library of metasurface lasers with photonic lattice systems in which one can design symmetries and degeneracies at will (e.g., honeycomb, Kagom\'{e} lattices, topological photonic crystals) \cite{weick_dirac-like_2013, proctor_higher-order_2021, zhang_observation_2018, heilmann_quasi-bic_2022}. Therefore, our work opens important new opportunities in the study of spontaneous symmetry breaking and emergence of spatial coherence. 

As example of the important role of symmetry and degeneracy, already in our hexagonal lattices one can compare $M$-point and $\mathit{K}$-point lasing \cite{juarez_m-point_2022}. For $\mathit{K}$-point lasing, the lattice symmetry rigorously guarantees that modes are exactly co-localized in space and exactly degenerate in energy. $M$-point lasing is also associated to six output spots, but instead arises from three distinct modes, each engaging two $M$-points. These modes are degenerate in frequency, but distinct in space. Consequently, in the $M$-point laser, we observe three-way spontaneous symmetry breaking that is easily turned into explicit symmetry breaking by changing the pump field polarization and thereby mode overlap with the near-field gain distribution. In this way, we can force lasing in just one out of the three modes. In contrast we were unable to coerce $\mathit{K}$-point lasing into either the $\mathit{K}$- or $\mathit{K'}$-mode by changing pump geometry, reflecting the exact degeneracy of the modes in space. 

Another important perspective is to study the spontaneous emergence of coherence in space. Our methods image spatial variations of the order parameters, a degree of freedom that is not present in simple coupled cavity realizations of SSB. In the reported experiment, we can for instance fit real-space patterns in boxes of just a few unit cells, and thereby assemble phase maps over the full microscope field of view (see supplement). In our system, such maps show mean-squared phase excursions of order 0.2 radians relative to the randomly chosen mean phase, with spatial correlation lengths that are of order $5$ $\mu$m. While the precise microscopic origin is out of the scope of this paper, these variations in part report on slight geometrical sample variations and in part arise from the distributed feedback lasing physics. 

According to Kogelnik and Shank \cite{kogelnik_coupledwave_1972}, even almost index-matched polymer distributed feedback laser show rich spatial structure that is determined by interplay of the system band structure, the balance between gain, loss, and overall laser size \cite{guo_spatial_2019}. It is completely open how this generalizes to the rich variety of nonlocal photonic systems that are currently emerging, encompassing plasmonic and dielectric metasurface lasers, photonic crystal lasers, quasi-Bound State in the Continuum, and topological lasers, as well as systems that show photon or exciton-polariton Bose-Einstein Condensate physics \cite{wertz_spontaneous_2010, kodigala_lasing_2017, ha_directional_2018, hakala_boseeinstein_2018}. Overall, our results demonstrate a practical route to quantifying emergence of coherence in  this entire array of photonic systems.

\section{Methods}
\paragraph{Sample fabrication}
We used  170 $\mu$m thin microscope cover slips (Menzel) as substrates. A 150 nm layer of PMMA was deposited on the substrate by spin-coating after cleaning them with base piranha. Then, we covered the PMMA with a thermally evaporated layer of 20 nm Ge, which will act as an etch mask for reactive ion etching after resist development. Over the Ge layer, we spin-coated a 65 nm layer of positive CSAR resist in which we wrote the structures using electron beam lithography (approximate dose of 130 $\mu$C/cm$^2$). The pattern was a hexagonal lattice of 500 nm period, consisting of 80 nm diameter particles. After exposure, we developed the samples for 60 s in pentyl-acetate, 6 s in O-xylene and finally 15 s in MIKB:IPA (9:1). We then etched through the Ge and PMMA layer by reactive ion etching. Finally, we evaporated a 30 nm Ag layer on the sample, and performed lift-off  in a 60 $^\circ$C acetone bath. This three-layer recipe from \cite{berkhout_perfect_2019} has the benefit of a high resolution, defined by the thin CSAR layer, and yet a thick resist stack with high undercut for the lift-off. After fabrication of the plasmon lattice, we spin-coated a 450 nm thick layer of SU8 doped with rhodamine 6G (10 mg Rh6G in 3 mL cyclopentanone mixed with 1 mL SU8, or 0.5 wt\% Rh6G) on the sample, as in \cite{schokker_lasing_2014}. 

\paragraph{Optical setup}
As the light source for pumping our plasmon lattice laser, we use the frequency-doubled output from a Light Conversion Pharos laser, providing 515 nm wavelength and 250 fs pulse duration. The electronic signal of the internal pulse picker is used to drive single-shot camera measurements. We direct the pump light through an epilens and a microscope objective (Nikon CFI Plan Apochromat lambda 100$\times$, NA=1.45), to obtain a 70 $\mu$m diameter collinear beam in the sample plane. The pump power is controlled by a motorized half-wave plate placed before a linear polarizer. The pump is filtered by a combination of a 532 nm dichroic mirror and a 550 nm longpass filter. We image the sample plane with the same objective onto a camera chip (Bassler acA1920-40um) for real-space imaging. A beam-splitter directs 50 \% of the light to another camera (Thorlabs CS2100M-USB) for Fourier-space imaging; in this track, we placed a Fourier lens in focus with the back focal plane of the objective via a 1:1 telescope. For spectrally resolved Fourier images (i.e, band structures), we direct the Fourier image through the entrance slit of a spectrometer (Andor Shamrock 163), to which we mounted a camera (Ximea MC124MG-SY-UB). For single-shot lasing experiments, we use a repetition rate of 20 Hz, and synchronize both the Bassler and Thorlabs camera to the laser, to enable simultaneous single-shot imaging in real-space and in Fourier-space. To obtain band diagrams in ($k_x, k_y$) space or in ($k_y, \omega$) space, we operate in multi-shot mode with a frame rate of 1 MHz and at low incident pulse power, thereby collecting below-threshold fluorescence. For the Fourier images below threshold, we insert a 550 nm (40 nm) bandpass filter, to select a narrow band of emission wavelengths near the lasing conditions.

\paragraph{$\mathit{K}$-point basis projection method}
We convert the sequence of images to a time series of integrated intensities of each of the 6 lasing output spots, obtained by summation of regions of interest (20 pixels across) around each spot. Having reduced the intensity information in the images to a vector of six intensities for each frame, we determine the projection of this 6-vector onto the subspace of lasing in superpositions of the $\mathit{K}$ and the $\mathit{K'}$ modes.  Let $I_N$ be the 6-vector of intensities in frame $N$ normalized to be of unit length, where the vector elements list intensities in the consecutive $\mathit{K}$-points enumerated in clockwise order, starting from the top spot. We determine the coefficients $c_{N, +}= \langle I_N | v_+\rangle $  and $c_{N,-}= \langle I_N | v_-\rangle $, where $v_\pm = 1\sqrt{6} (1,\pm 1, 1,\pm 1, 1, \pm 1)$, and $\langle . | \rangle$ stands for the inner product. We normalize to $\langle I_N|I_N\rangle$.  The normalized temporal autocorrelation for a parameter $x$ is defined as 
$\mathrm{Mean}[(x(n+\Delta n) -\bar{x})(x(n) -\bar{x})]/\mathrm{Var}[x]$ where $\Delta n$ refers to comparing the time series at $\Delta n$ shots apart. 

\paragraph{Spatial filter and relative phase fit methods}
See Supporting Information.

\newpage
\bibliography{DeGaayFortmanKoenderink}
\newpage

 \setcounter{figure}{0}
\renewcommand{\thefigure}{S\arabic{figure}}

\section*{Supplement to:  Spontaneous symmetry breaking in plasmon lattice lasers}
\setcounter{section}{0}
\section{Spectrum of the lasing mode}
\begin{figure*}
    \centering
    \includegraphics[width=1\textwidth]{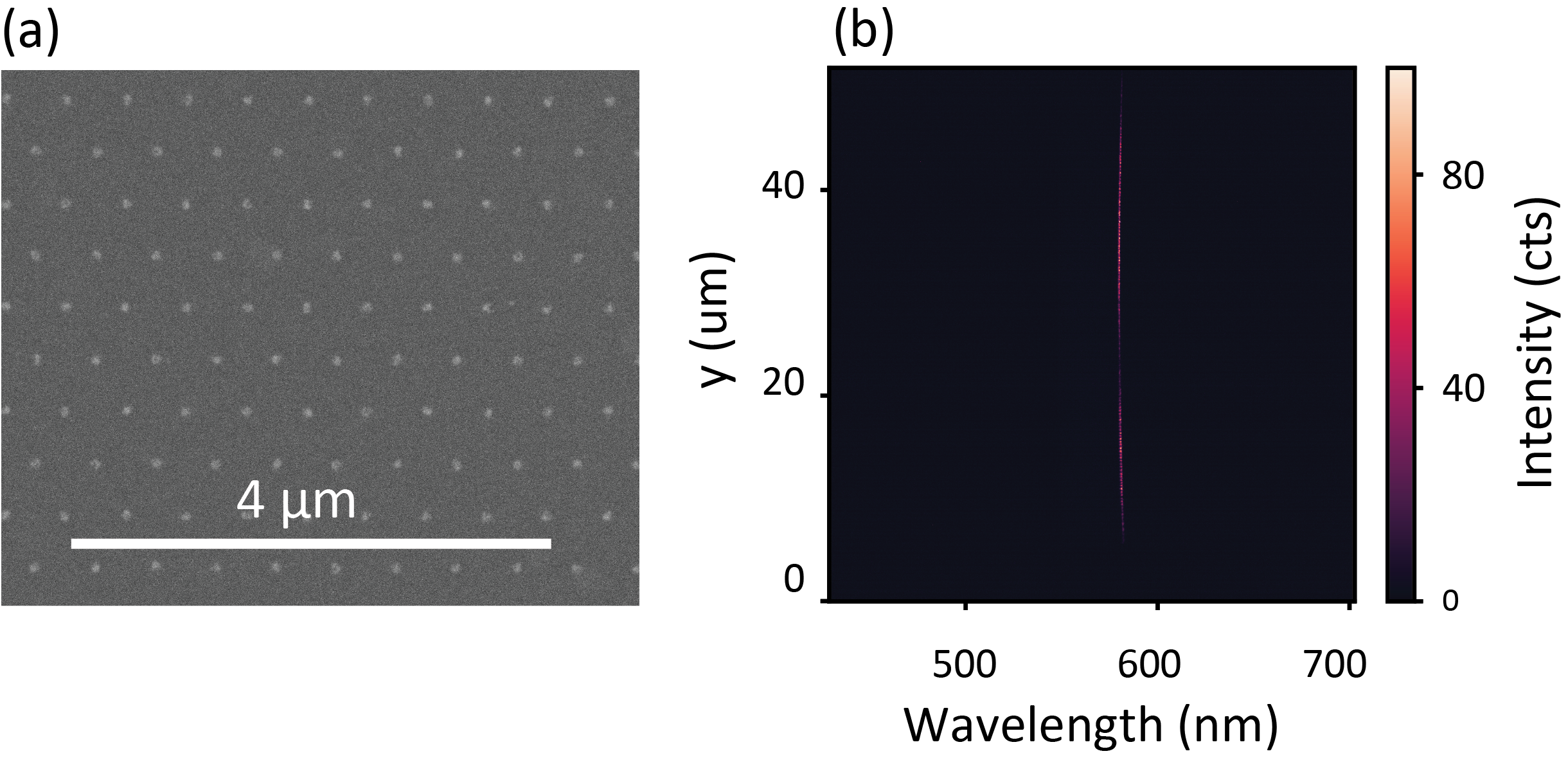}
    \caption{(a) SEM image of the $\mathit{K}$-point plasmonic lattice laser, showing the lattice of pitch 500 nm of silver plasmon particles with diameter 80 nm. (b) raw spectral image of $\mathit{K}$-point lasing emission, showing a single lasing line along the full length of the real-space field of view. The slight curvature is due to the spectrometer imaging optics and has not been corrected for in this image.}
\end{figure*}

The $\mathit{K}$ and $\mathit{K'}$ modes both form a set of 3 k-space points that are coupled by a reciprocal lattice vector. There exist no lattice vectors that could couple $\mathit{K}$ into $\mathit{K'}$ and vice versa, which ensures the energy degeneracy between modes that live at the $\mathit{K}$/$\mathit{K'}$-points. In essence, the energy degeneracy is thus a result of Bloch's Theorem. Fig S1(b) further evidences this degeneracy. While Fig 2(f) in the main manuscript shows the spectra taken in Fourier space, separating out $\mathit{K}$ and $\mathit{K'}$, Fig S1(b)  shows  spatially resolved spectroscopy, obtained by projecting a slice of the real space image onto the spectrometer slit. The spectral image is not corrected for imaging artifacts in the Andor spectrometer, and therefore there is a slight curvature in the result. Along the entire length of the field of view that is imaged (circa 44 $\mu$m) there is only a single lasing line.

\section{SSB power dependence}
\begin{figure*}
    \centering
    \includegraphics[width=0.75\textwidth]{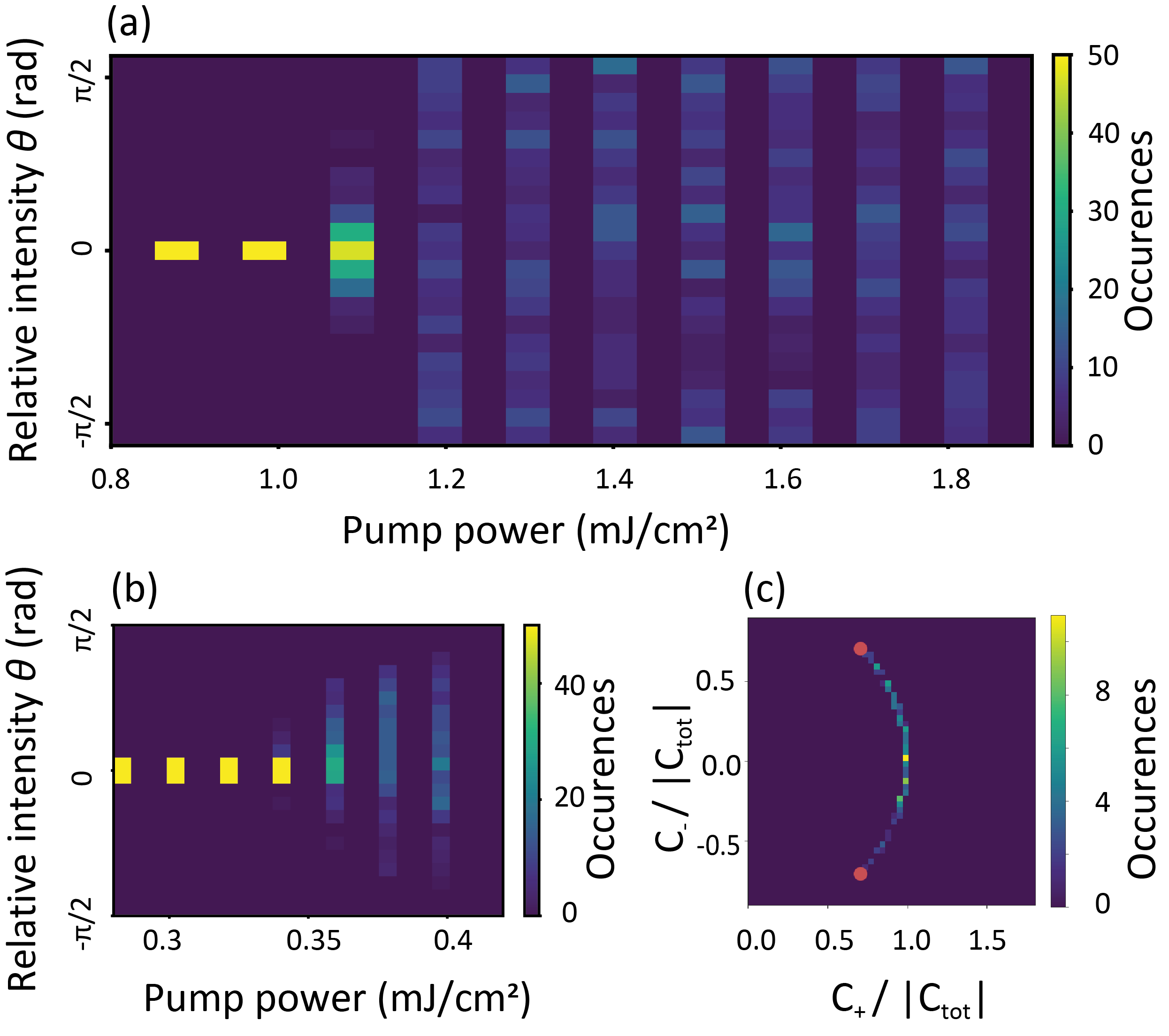}
    \caption{(a) Histograms of $\mathit{K}$/$\mathit{K'}$ intensity contrast parameter $\theta$ for increasing pump fluence, showing even spreads in $\theta$ for fluences well above lasing threshold. Data taken with the 250 fs pulsed laser setup described in the main text. (b) Same measurement as in (a) but taken with a setup using a laser of 500 ps pulse length. (c) Fluorescence intensity background subtracted occurrence histogram for $\theta$, showing that the fluorescence background significantly reduces the spread in $\theta$.}
\end{figure*}

Spontaneous symmetry breaking generally occurs when one input parameter to a physical system is continuously changed, and crossing some threshold value, the system's energy mode reduces in symmetry. Fig S2(a) displays such threshold behavior for the $\mathit{K}$/$\mathit{K'}$ mode laser, mapped as function of pump power. We use the parameter $\theta$ defined in the main text. Below lasing threshold, fluorescence at all $K$-points is equally bright ($\theta=0$). As the pump power exceeds the lasing threshold power of 1.0 mJ/cm$^{2}$, a continuous spread becomes visible in the SSB parameter $\theta$ that expresses the relative intensity between the $\mathit{K}$ and $\mathit{K'}$ mode ($y$-axis). The spread remains uniform over the entire  $\theta$ range  when increasing pump power. Fig S2(a) is measured with the 250 fs pump pulses also used for the main manuscript.  For Fig S2(b)  we placed the same sample in a similar fluorescence Fourier microscope but with 532 nm pump pulses of 500 ps pulse length, and we applied the same linear basis projection method to the Fourier images. Also for this much longer pump pulse regime, the system shows no bifurcation, and a wide range of $\theta$ values is sampled. The spread in $\theta$ is less than in Fig S2(a). We attribute this to a higher fluorescence intensity background when using 500 ps long pulses than 250 fs, which reduces $\mathit{K}$/$\mathit{K'}$ contrast. This is supported by Fig S2(c) that shows the histogram for 0.4 mJ/cm$^{2}$, but with fluorescence background subtracted prior to extracting $\theta$. This recovers a more uniform distribution of $\theta$ over the full range.

\section{Coupled-mode model}
\begin{figure*}
    \centering
    \includegraphics[width=0.75\textwidth]{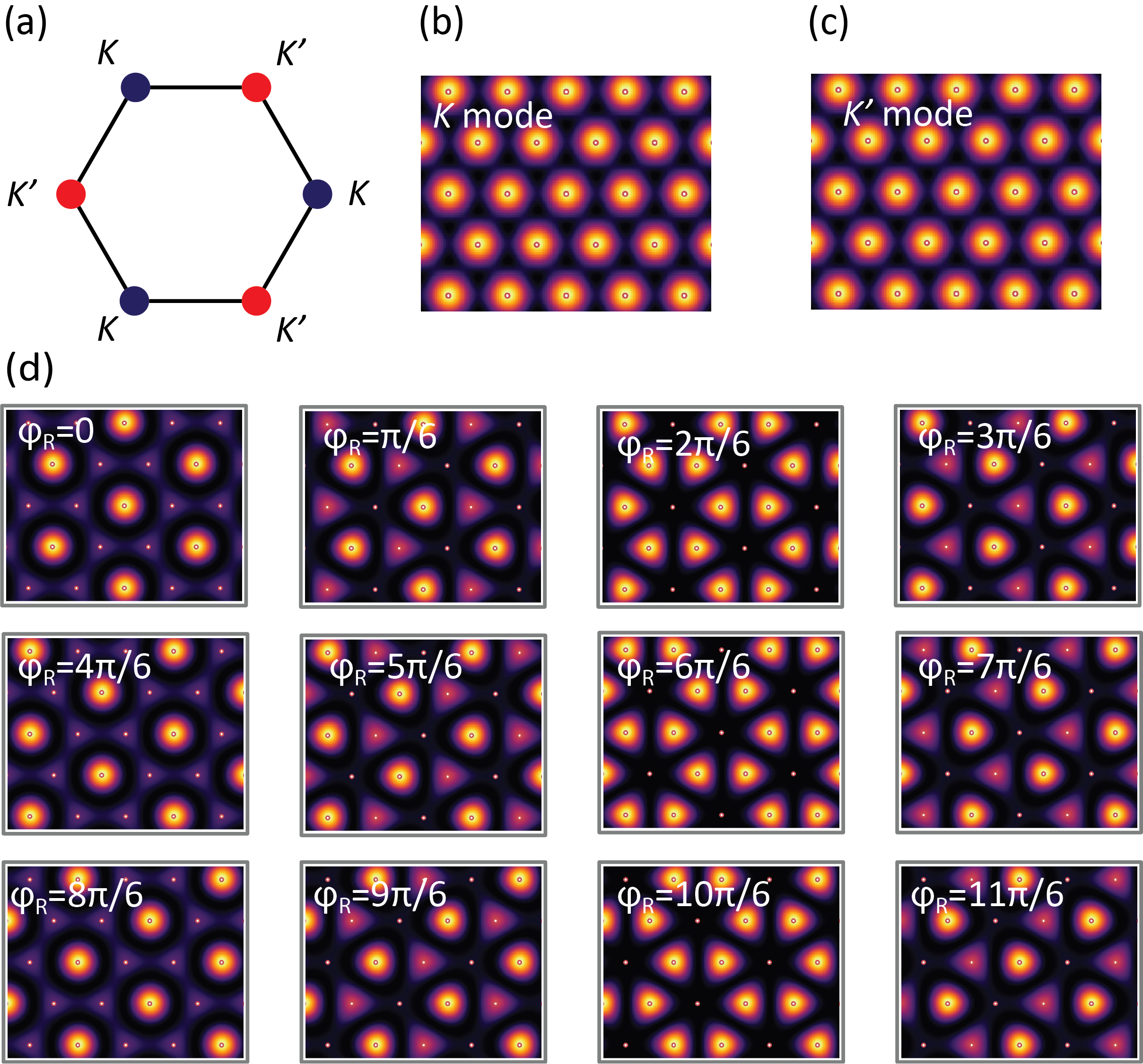}
    \caption{(a) The hexagonal Brillouin zone corners comprise two sets of three reciprocal lattice vector-coupled $\mathit{K}$-points. The $A$-type modes of the $\mathit{K}$/$\mathit{K'}$ points are calculated with a coupled-wave model, they have equal intensity distributions in space, as shown in (b) ($\mathit{K}$ mode) and (c) ($\mathit{K'}$ mode). (d) Catalogue of local $|\mathbf{E}_{T}|^2$ for different relative phases $\varphi_R$ between $\mathit{K}$ and $\mathit{K'}$ modes (calculated with equal $\mathit{K}$-mode amplitudes $a=b$).}
\end{figure*}

From scalar coupled-mode theory, the $A_1$ $\mathit{K}$/$\mathit{K'}$-point Bloch mode is described by three coupled waves (see Fig S3(a)): 
\begin{equation}
    \mathbf{E}_{\mathbf{\mathit{K}}} = e^{i\mathbf{\mathit{K}_1}\cdot\mathbf{r}} + e^{i\mathbf{\mathit{K}_2}\cdot\mathbf{r}} + e^{i\mathbf{\mathit{K}_3}\cdot\mathbf{r}}
\end{equation}

This simple model is used to calculate local field distributions for the $A_1$ mode. Figs S3b,c display the calculated intensity distributions of the $\mathit{K}$/$\mathit{K'}$ modes, showing that they both form hexagonal patterns with peaks at the lattice particle positions. The $\mathit{K}$/$\mathit{K'}$ modes are thus spatially degenerate and are only distinguished by their opposite wavevector content. When both modes form a coherent superposition
\begin{equation}
    \mathbf{E}_{T} = a \mathbf{E}_{\mathit{K}} + b \mathbf{E}_{\mathit{K'}}e^{i\varphi_R}
\end{equation}
the result in local intensity shows honeycomb interference patterns as visible in Fig S3(d). The supercell pattern has $\sqrt{3}$ times larger pitch than the sublattice, and the structure and spatial alignment relative to the origin of the patterns vary with relative phase $\varphi_R$. 

\section{Real space filtering and phase fitting}
\begin{figure*}
    \centering
    \includegraphics[width=0.75\textwidth]{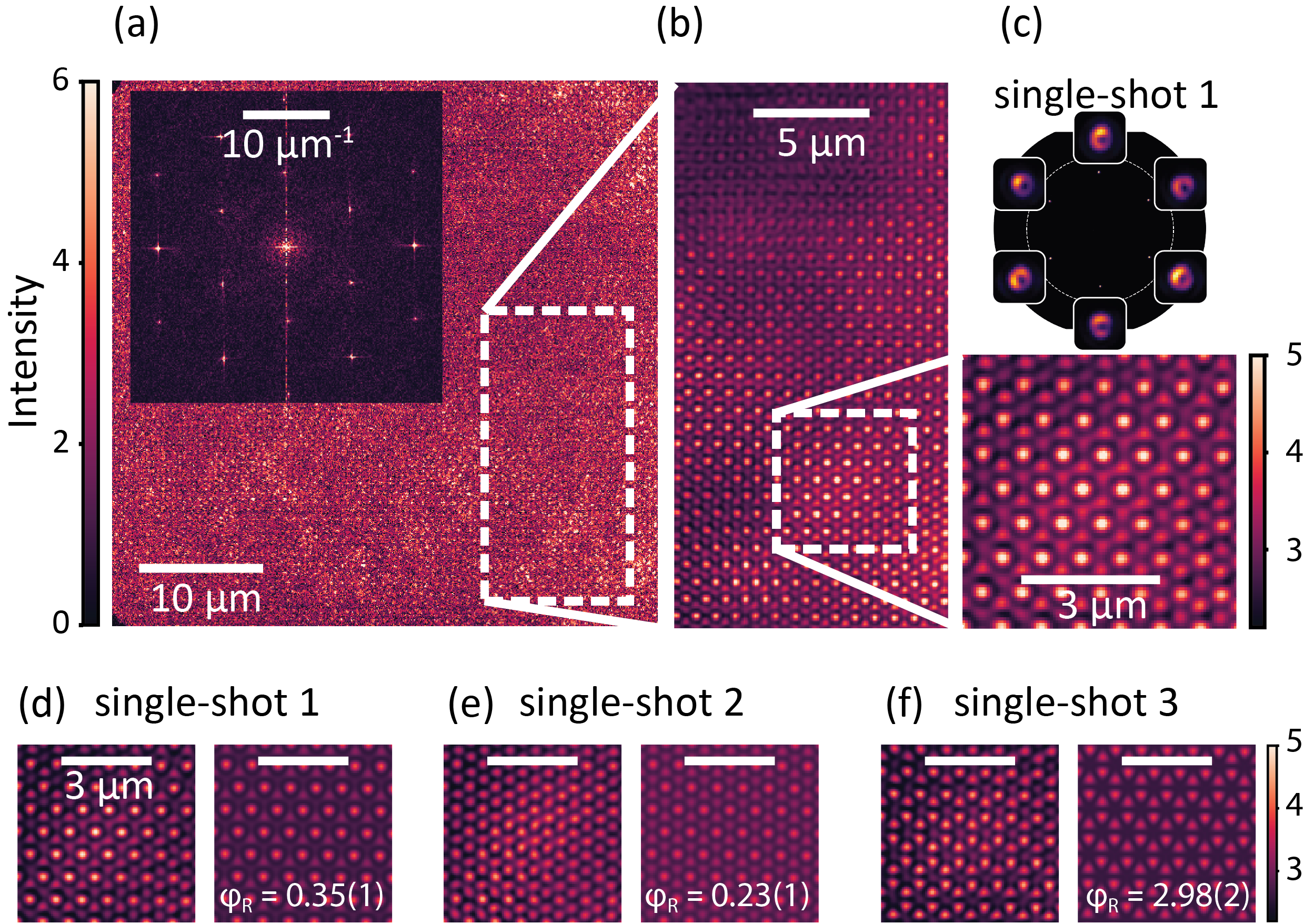}
    \caption{Workflow illustration of the method to extract relative phase between the $\mathit{K}$ and $\mathit{K'}$ mode from the real space data. (a) Raw image and FFT as inset. (b) Fourier filtered data for region as indicated in (a). (c) Close up of area used for fitting, taken from (b). Also shown is the corresponding measured Fourier image. Panels (d-e) show example data and fits.}
\end{figure*}

In a single-shot real space image (Fig S4(a)), different periodic features are distinguishable, but their detailed structures are partially obscured by residual noise from scattering by fabrication disorder. We Fourier filter the real space data for fitting, as follows. The inset of Fig S4(a) shows the absolute value of the 2D Fourier transform $F$ of the real space image, in which we clearly observe 19 sharp peaks (one at 0 wave vector, and the remaining 18 forming different orders spanning the honeycomb patterns hexagonal lattice). In the 2D FFT peaks, 6 points at $|k//| = 15$ $\mu$m$^{-1}$ are the first order reciprocal lattice points of the real space lattice itself, and the remaining 12 are responsible for the superlattice periodicity. For fitting, we first define the size $l=5.86$ $\mu$m of the real space blocks of data that we will fit the coupled-mode model to. This defines a Fourier filter width $dk=2\pi/l$ $\mu$m$^{-1}$. We then remove high-frequency noise from the full real space image by selecting only regions of width $\Delta k$ around each of the 19 sharp peaks (Gaussian filters, width $dk$). Inverse-Fourier transforming the masked FFT returns the filtered real space image. Intensities in real-space images in Fig S4 are reported in dimensionless units. For a quantitative scale, the data in Fig S4(a) corresponds to $1.4 \cdot 10^6$ camera ADU units integrated over the 44 by 44 $\mu$m$^2$ field of view. With a camera quantum efficiency of 70\% this translates into circa $1 \cdot 10^6$ collected photons per laser shot (500 photons per square micron). The Fourier filtering suppresses the Poisson counting noise.

We then fit filtered real space intensity patterns of size $l\times l$ to the coupled-mode model with the Scipy minimize function using the BFGS algorithm. For all the fits, we choose the same area in the sample, and the block origin is chosen to coincide with an intensity peak in the ensemble-averaged data. This ensures a precise and common registration to the underlying particle lattice, which is needed to avoid ambiguity in fitting the phase. The fit function reads:

\begin{equation}
    |\mathbf{E}_{T}|^2 = A|(\sqrt{a} \mathbf{E}_{\mathit{K}} + \sqrt{b} \mathbf{E}_{\mathit{K'}}e^{i\varphi_R})|^2 + B
\end{equation}

where the relative amplitude coefficients $a$ and $b$ are derived from $\theta$, taken from the Fourier-space measurement, as: $a = (1+\tan{\theta/2})/2$ and $b = (1-\tan{\theta/2})/2$. The only free fit parameters are the overall amplitude and background counts ($A$ and $B$) and the relative phase $\varphi_R$.  Representative fit results are shown in Figs S4(d,e,f). We note that performing this fit in real space works best at choices of $l$ that encompass of order 10x10 unit cells.

\section{Spatial landscapes of the phase}
\begin{figure*}
    \centering
    \includegraphics[width=0.75\textwidth]{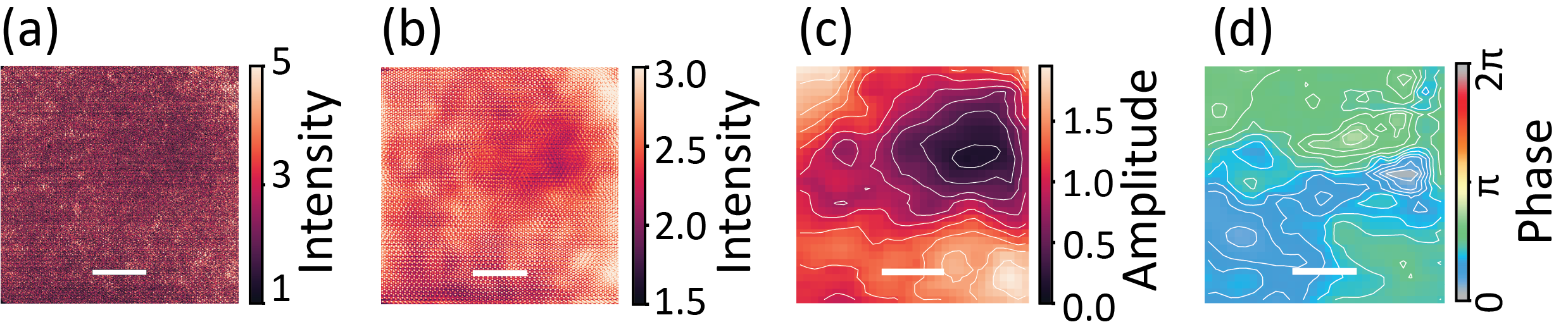}
    \caption{Spatial phase landscape of a single lasing shot. (a) The raw single-shot real space data, to which the spatial filter technique is applied to obtain (b). (c) Fitted amplitude of the periodic patterns above the fluorescence background. (d) Spatial landscape of the extracted relative phase between $K$ and $K'$ mode. Scale bars are 10 um. }
\end{figure*}

The fitting procedure as described above can be extended to obtain a spatial phase texture over the microscope field of view. To this end, we dive the field of view in  a target set of phase sampling points and for each sampling point we determine the phase by real space fitting the intensity pattern inside a small box centered at the sampling point (side length 5$\times$ larger than the sampling point distance). For successful fitting, it is important that in all images the underlying particle lattice is precisely referred to one common origin, since otherwise an error in spatial position would translate in a phase slip (see Fig S3).  We have found it advantageous to first fit the phase texture at a coarse sampling resolution, fitting larger areas a time, and using those fit results as initial guess for fitting at refined resolution. Generally, we find that the local phase can be fitted accurately except in two cases. First, whenever the lasing is purely $\mathit{K}$ or $\mathit{K'}$, the relative phase is $\varphi_R$ not defined. In practice, this means that also close to pure $\mathit{K}$/$\mathit{K'}$-lasing the phase is hard to determine.  Second, the amplitude maps show variations in laser intensity, including marked minima.  We note that such inhomogeneities in amplitude envelopes are intrinsic to DFB lasing,  and for instance occur in coupled mode theory descriptions from a nonlinear interplay of laser size, gain coefficient and feedback strength in coupled mode theory. At intensity minima the contrast in the data to which the fit is performed is low. Fig S5 shows the results for one example frame, reporting the raw real space image, the Fourier filtered data, and the extracted phase map. For this particular example there are phase excursions of order 0.1 to 0.2 radians around the mean. Fig S6 shows such phasemaps for 6 subsequent frames. Both the average phase, and the phase fluctuations vary from map to map. The spatial autocorrelation lengths are of order 5 $\mu$m.

\begin{figure*}
    \centering
    \includegraphics[width=1\textwidth]{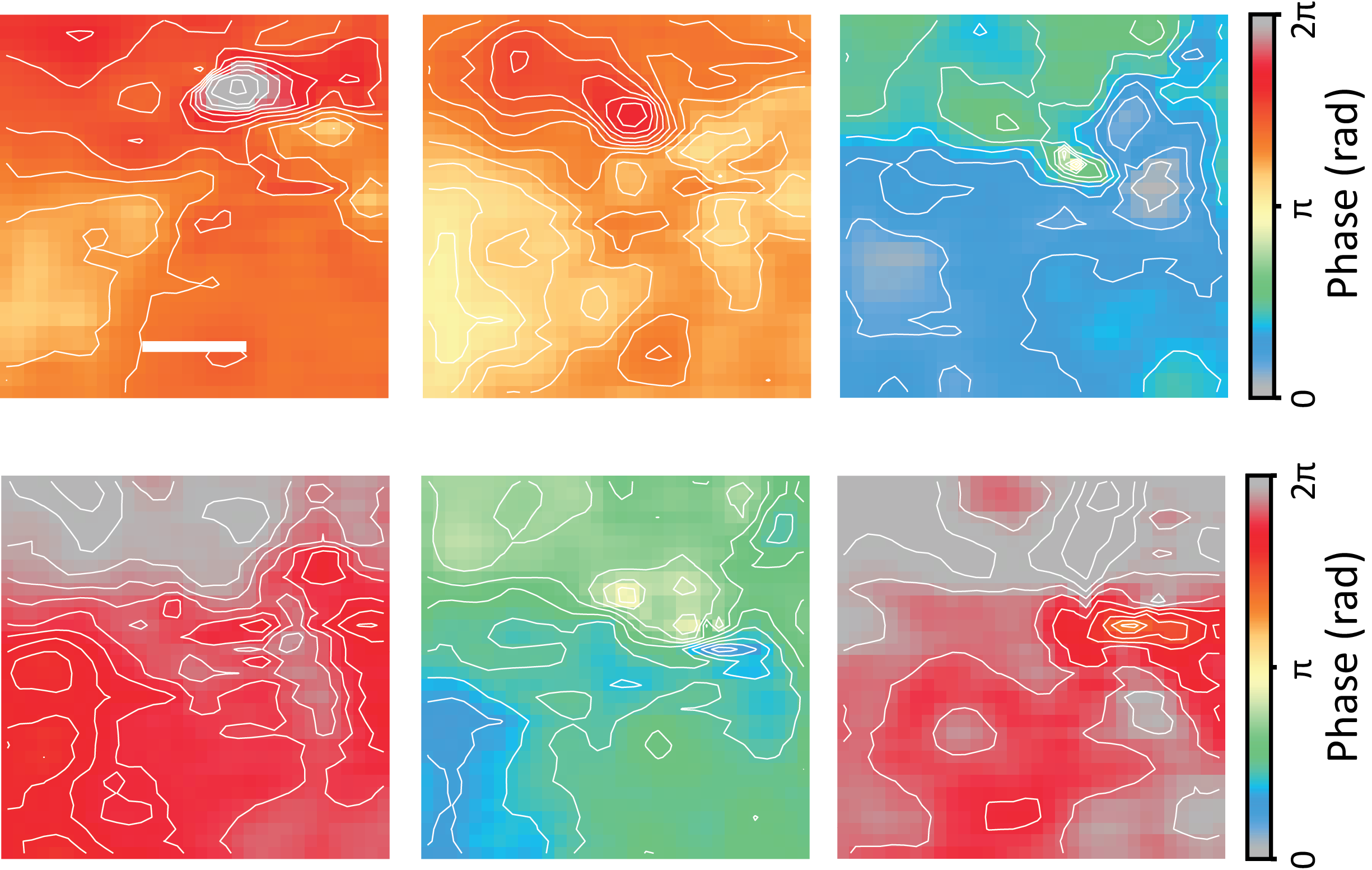}
    \caption{Spatial phase landscapes for six subsequent single-shots, showing random excursions from the mean spatial phase texture.}
\end{figure*}

%\end{document}

\end{document}